\pdfoutput=1

\documentclass[sigconf, nonacm]{acmart}

\usepackage{xspace}

\newcommand\system{{\sf QStore}\xspace}
\newcommand\format{{\sf QStore}\xspace} 

\newcommand\safetensors{{\sf Safetensors}\xspace}

\newcommand\lzfour{{\sf lz4}\xspace}
\newcommand\zstd{{\sf Zstd}\xspace}
\newcommand\zipnn{{\sf ZipNN}\xspace}

\usepackage{subcaption}
\usepackage{pgfplots}
\usepackage{pgfplotstable}
\usepackage[table]{xcolor}
\usepackage{tikz}
\usepackage[capitalise]{cleveref}
\usetikzlibrary{calc}
\usetikzlibrary{fit}
\usetikzlibrary{patterns}
\usetikzlibrary{decorations.pathreplacing}
\usetikzlibrary{arrows,calc,arrows.meta}
\usetikzlibrary{matrix,positioning}
\usetikzlibrary{shapes.geometric}

\definecolor{BaseBothColor}{HTML}{264653}
\definecolor{Safetensors}{HTML}{264653}
\definecolor{BaseLargeColor}{HTML}{777777}
\definecolor{UncompressedColor}{HTML}{777777}
\definecolor{LzfourColor}{HTML}{2a9d8f}
\definecolor{ZstdColor}{HTML}{e9c46a}
\definecolor{ZipnnColor}{HTML}{e76f51}
\definecolor{OursColor}{HTML}{C62E2E}

\definecolor{BlueColor}{HTML}{0081a7}
\definecolor{GreenColor}{HTML}{38A528}
\definecolor{Lightgrey}{HTML}{dadada}

\definecolor{color1}{HTML}{7081ff}
\definecolor{color2}{HTML}{ffb0c2}
\definecolor{color3}{HTML}{ABDDA4}
\definecolor{color4}{HTML}{38A528}

\crefname{section}{§}{§§}
\Crefname{section}{§}{§§}
\crefformat{section}{\S#2#1#3}
\crefname{figure}{Fig}{Figs}
\Crefname{figure}{Fig}{Figs}
\crefname{problem}{Prob}{Probs}
\Crefname{problem}{Prob}{Probs}

\newcommand\vldbdoi{XX.XX/XXX.XX}

\newcommand\vldbavailabilityurl{https://github.com/illinoisdata/qstore}

\begin{document}
\title{
QStore: Quantization‑Aware Compressed Model Storage 
}

\author{Raunak Shah}
\affiliation{%
  \institution{Univ. of Illinois Urbana-Champaign}
}
\email{raunaks3@illinois.edu}

\author{Zhaoheng Li}
\affiliation{%
  \institution{Univ. of Illinois Urbana-Champaign}
}
\email{zl20@illinois.edu}

\author{Yongjoo Park}
\affiliation{%
  \institution{Univ. of Illinois Urbana-Champaign}
}
\email{yongjoo@illinois.edu}

\begin{abstract}
Modern applications commonly leverage large, multi-modal foundation models, in complex workflows that demand the storage and usage of similar models in multiple precisions.
A straightforward approach is to maintain a separate file
    for each model precision (e.g., INT8, BF16),
which is indeed taken by model providers
    such as HuggingFace and Ollama.
However, this approach incurs excessive storage costs
    as a higher precision model (e.g., BF16)
    is a superset of a lower precision model
        (e.g., INT8) in terms of information.
Unfortunately,
    simply maintaining only the higher-precision model
    and requiring every user to dynamically convert the model precision is not desirable
because every user of lower precision models
    must pay the cost for model download and precision conversion.

In this paper, we present QStore, a unified, lossless compression format for simultaneously storing a model in two (high and low) precisions efficiently. 
Instead of storing low and high-precision models separately,
\format stores low-precision model and only \emph{residual information} needed to reconstruct high-precision models.
The residual information size is significantly
    smaller than the original high-precision models,
    thus, achieving high storage cost savings.
Moreover, \format does not compromise model loading speed:
The low-precision models can still be loaded quickly, while the high-precision models can also be reconstructed efficiently by merging low-precision data and the residual with \format's lightweight decoding.
We evaluate \format for compressing multiple precisions of popular foundation models, and show that \format reduces overall storage cost by up to $2.2\times$ while enabling up to $1.7\times$ and $1.8\times$ faster model saving and loading versus existing approaches.

\end{abstract}

\maketitle



\newcommand{\midsepremove}{\aboverulesep = 0.3mm \belowrulesep = 0.3mm}
    \midsepremove
    \newcommand{\midsepdefault}{\aboverulesep = 0.605mm \belowrulesep = 0.984mm}
    \midsepdefault

\section{Introduction}
\label{sec:intro}
Foundation models have become highly accessible to users thanks to the availability of model hosting platforms such as HuggingFace~\cite{huggingface-transformers}, Ollama~\cite{ollama}, and ModelScope~\cite{modelscope}. Developers download pre-trained models hosted on these platforms (e.g., from cloud storage), and then apply them to tasks such as fine-tuning~\cite{fine-tuning, eacl-fine-tuning, riskaverse-fine-tuning}, distillation~\cite{distillation, autoreg-knowledge-distillation} and inference~\cite{inferencesurvey, 8bitinference, wu2025cache, zhang2150pbe, fang2025large}. 
Different tasks demand different model precisions; for example, fine-tuning is often performed using higher precisions such as FP16~\cite{fp16-fine-tuning}, then, the fine-tuned model would be \textit{quantized} to a lower precision such as INT8~\cite{compevalquantization, databricksblog} for faster inference. Hence, many workflows require accessing the same model under different precisions: in addition to fine-tuning-then-inference, other tasks with this requirement include Model Cascade~\cite{chen2023frugalgpt, zhang2024efficient} and Chaining~\cite{wu2022ai, wei2022chain, besta2024graph}. Moreover, data scientists and researchers also iterate between different-precision models for testing and benchmarking~\cite{nvidiablog, reddit_user}. 


\begin{figure}[t]
\begin{subfigure}{\columnwidth}
\centering
\begin{tikzpicture}[>={LaTeX[width=1mm,length=1mm]},->]
\tikzset{cbox/.style={
	minimum height=10mm,draw=black,fill=white,ultra thick,align=center,
	minimum width=14mm,anchor=north
}}
\tikzset{dbox/.style={
	minimum height=6mm,minimum width=10mm,draw=black,fill=white,thick,
    font=\scriptsize, align=center,anchor=north,inner sep=0.5mm
}}

\tikzset{
mylabel/.style={
    font=\footnotesize\sffamily\bfseries,
    align=center,
},
mylabel2/.style={
    font=\footnotesize\sffamily,
    align=center,
},
mycomponent/.style={
    semithick, rounded corners=0.5mm,
}
}

\node(inputs1) [draw=black, fill = Lightgrey, mycomponent,align=center,font=\sffamily\bfseries,anchor=east, minimum width = 13mm] at (0, 0) {0100111011010000};
\node(inputs2) [draw=black, fill = Lightgrey, mycomponent,align=center,font=\sffamily\bfseries,anchor=north, minimum width = 13mm] at ($(inputs1.south)$) {0100111011010011};
\node [anchor=east, align=center,font=\sffamily\bfseries] at ($(inputs1.west)$) {27.25};
\node [anchor=east, align=center,font=\sffamily\bfseries] at ($(inputs2.west)$) {27.30};
\node [anchor=south, align=center,font=\sffamily\bfseries] at ($(inputs1.north) + (0, 0)$) {High-precision model weights (FP16)};

\node(quantize) [draw=black, mycomponent,align=center,font=\sffamily\bfseries,anchor=north, minimum width = 13mm] at ($(inputs2.south) + (0, -0.3)$) {Quantization\\[-0.2em]\footnotesize (rounding, scaling, etc.)};

\node(inputs3) [draw=black, fill = Lightgrey, mycomponent,align=center,font=\sffamily\bfseries,anchor=north, minimum width = 13mm] at ($(quantize.south) + (0, -0.3)$) {01100110};
\node [anchor=east, align=center,font=\sffamily\bfseries] at ($(inputs3.west)$) {102};
\node [anchor=north, align=center,font=\sffamily\bfseries] at ($(inputs3.south) + (0, 0)$) {Common Information (INT8)};

\node(inputs5) [draw=black, fill = Lightgrey, mycomponent,align=center,font=\sffamily\bfseries,anchor=south west] at ($(inputs3.east) + (2.0, 0)$) {01};
\node(inputs6) [draw=black, fill = Lightgrey, mycomponent,align=center,font=\sffamily\bfseries,anchor=north] at ($(inputs5.south) + (0, 0)$) {10};

\draw[->, thick] 
($(inputs2.south) + (0, 0)$) --
($(quantize.north) + (0, -0)$); 
\draw[->, thick] 
($(quantize.south) + (0, 0)$) --
($(inputs3.north) + (0, -0)$); 
\draw[->, thick] 
($(inputs3.east) + (0, 0)$) --
($(inputs5.south west) + (0, -0)$); 

\draw[-, thick] 
($(inputs1.south east) + (0, 0)$) --
($(inputs1.south east) + (1.65, -0)$); 

\draw[->, thick] 
($(inputs1.south east) + (1.65, -0)$) --
($(inputs5.north) + (0, -0)$); 

\node [anchor=west, align=left,font=\sffamily\bfseries] at ($(inputs5.south east) + (0, 0)$) {Conditional\\[-0.2em] information\\[-0.2em] (FP16|INT8)};
\end{tikzpicture}
\end{subfigure}
\caption{\format stores the conditional bit representation of the high-precision model weights alongside the common low-precision, quantized weights to store a model in both high and low-precisions using fewer bits.}
\label{fig:intro}
\end{figure}

\paragraph{Storing Multiple Models is Costly}
Currently, a common approach to maintaining multiple models of varying precisions while doing the aforementioned tasks is to store them as is (i.e., separately storing the multiple precision versions)~\cite{nvidiablog, reddit_user}. However, as newer, more complex tasks demand ever-increasing model sizes (e.g., Mistral-7B~\cite{jiang2023mistral7b} being sufficient for simple math tasks, while more complex, multi-modal tasks~\cite{wu2025cache} require larger models such as Qwen2.5-VL 32B~\cite{Qwen2.5-VL}), the storage cost incurred by storing multiple model versions can become prohibitive --- for example, 91.8 GB is required to store just the BF16~\cite{bf16study} and INT8 (quantized) versions of the Deepseek-Coder~\cite{deepseek-coder} 33B parameter model. The space cost is a significant issue for developers using these models, and also increases cloud storage costs for model hubs like HuggingFace, Ollama, and ModelScope, since model providers end up storing multiple precisions of these models separately on these platforms to account for user accesses to models in different precisions.

One potential approach to reduce storage cost is to only store the highest-precision model (e.g., FP16 or BF16), then quantize in-memory if lower precisions (e.g., INT8) are needed~\cite{gptq}. However, retrieving a low-precision model with this approach is inefficient as it requires (i) loading more data than necessary (i.e., the high-precision model) and (ii) an expensive quantization process (e.g., up to 21 GPU minutes for a 13B model~\cite{gptq}). Alternatively, models can be compressed with an algorithm such as LZ4~\cite{lz4}, ZSTD~\cite{zstd}, or ZipNN~\cite{zipnn}. However, these algorithms either use generic techniques that underperform on ML weights (e.g., LZ4 and ZSTD), or are tailored to specific precisions (e.g., ZipNN for FP16/BF16).

\paragraph{Our Intuition}
We propose \format, a data format for efficiently storing varying precision versions of a model. We observe that despite being quantized, a lower-precision (e.g., INT8) model version contains information that is also present in a higher-precision (e.g., FP16, BF16) version. Hence, compared to separately compressing and storing a pair of higher and lower-precision models, it is possible to use less space to \textit{simultaneously} represent both models. \Cref{fig:intro} illustrates this idea: much of the information in the weights of a high-precision FP16 model is already contained in the low-precision (i.e., quantized) INT8 version. 
Hence, given an already efficiently stored low-precision model, we can also store the high-precision model using only a few \textit{additional} bits per weight representing `extra information' not in the low-precision model (i.e., the `FP16 | INT8' \textit{conditional} model). Such a unified data format would (1) save storage space versus storing both models separately (regardless of compression), (2) enable faster loading of the lower-precision model versus loading a high-precision model and quantizing it, while (3) still enabling fast loading of the high-precision model.

\paragraph{Challenges}
Designing a unified data format for simultaneously and efficiently storing a pair of high and low-precision models is challenging. 
First, we need to carefully define `extra information' not present in the lower-precision model required to reconstruct the higher-precision model. Significant information is lost while quantizing a higher-precision model to a lower-precision one (e.g., from operations like rounding), hence, our definition should effectively encapsulate this information gap for lossless reconstruction. Identifying this information gap is nontrivial, as a quantized weight may be significantly different from the original weight in both bit representation and numerical magnitude (\cref{fig:intro}). Second, our representations of the lower-precision model's information and `extra information' should strike an acceptable storage/processing speed trade-off: for example, na\"ively defining and storing information at a bit-level granularity would enable the most efficient model storage, but can result in unacceptable model loading and saving times.




\paragraph{Our Approach}  

Our key idea for \format is to design a generalized compressed representation for conditional information that can work well despite the differences between floats and integers; such a format would allow us to load low and high-precision models, regardless of their data type, with perfect accuracy.



First, for storage, given a high and low-precision model pair's weights, we separately encode the low-precision model weights and \textit{conditional} weights (i.e., the 'extra information') with novel entropy coding and intelligent grouping strategies to enable significantly better compression ratios versus separately compressing the two models using off-the-shelf compression algorithms.

Then, for model loading from \format, we process the encoded low-precision model's weights, or additionally the conditional weights, to retrieve the low-precision or high-precision model, respectively. We perform decoding at a byte-level granularity to ensure high decoding speeds on common computing architectures~\cite{bytebitgranularity}. Our decoding is notably lossless (e.g., versus dequantization~\cite{dequantize}).

\paragraph{Contributions}
Our contributions are as follows:
\begin{enumerate}
    \item \textbf{Format.} We describe how \format, a data format to efficiently store a high and low-precision model pair. (\cref{sec:system})

    \item \textbf{Usage.} We describe efficient encoding and decoding schemes for storing/loading models to/from \format. (\cref{sec:compression})
    \item \textbf{Evaluation.} 
    We verify on 6 foundation models that \format reduces model pair storage costs by up to $55\%$, enables up to $1.6\times$ and $2.2\times$ faster loading and saving, respectively, versus alternative methods, and generalizes to various data types, quantization algorithms, and >2 model chains. (\cref{sec:experiments})
    

\end{enumerate}
\section{Background}
\label{sec:motivation}
Efficiently storing and deploying large foundation models is challenging. Our work addresses this challenge through proposing a compressed format capable of concurrently storing multiple model representations of different precisions. This section covers related work on quantization (\cref{sec:motivation_quantization}) and compression (\cref{sec:motivation_compression}).




\subsection{Quantization}
\label{sec:motivation_quantization}
Quantization is commonly applied to models to achieve desired quality-resource consumption tradeoffs. In this section, we overview the pros and cons of common quantization techniques, and key differences between \format and quantization.

\paragraph{Common Quantization Targets}
While 32-bit floating-point (FP32) precision was once standard~\cite{mixedprectraining}, the recent increases in model sizes and corresponding increases in computational and memory requirements have driven the adoption of lower-precision, quantized model formats.
For example, 16-bit precision (FP16~\cite{fp16_nvidia}, BF16~\cite{bf16_google, bf16study}) formats have become a de-facto standard for training and fine-tuning to balance between accuracy and resource consumption.
For more resource-constrained scenarios or latency-sensitive applications (e.g., on-device processing~\cite{ondevicellms}), further quantization is common---typically to 8-bit (INT8) ~\cite{jacob2018quantization, llmint8dettmers}, but sometimes more aggressively to 4-bit (INT4, NF4) ~\cite{dettmers2023qlora, gptq, awq} or even lower~\cite{BitNet}. Recently, FP8 quantization has also been used during inference~\cite{fp8}.

\paragraph{Quantization Methods}
There exists several notable classes of quantization methods commonly applied to foundation models. (1) \textbf{RTN (round to nearest)} rounds weights to the nearest representable value in low-precision format (e.g., $42.25 \rightarrow 42$), which is fast, but can significantly degrade model accuracy (e.g., with outlier weights). (2) \textbf{Channel-wise quantization} such as LLM.int8()~\cite{llmint8dettmers} and SmoothQuant~\cite{smoothquant} apply per-channel scaling and quantization to model weights to better preserve outliers. (3) \textbf{Reconstruction-based approaches} such as AWQ~\cite{awq} and GPTQ~\cite{gptq} are also applied on a per-channel or per-block level, but they aim to quantize in a fashion such that the original high-precision weights can be reconstructed with minimal error. While these methods are capable of quantizing to very low precisions such as INT4 and INT3, they incur higher computational overhead versus alternatives.

Quantization methods operate at a per-block level, since it allows them to be efficient, permitting parallelization over multiple threads (including GPUs), and requiring less metadata compared to quantizing every element separately. We will later show how \system utilizes this standardized block-based approach to be generally applicable to various quantization methods (\cref{sec:compression_encoding}).




\paragraph{\format vs Lossy Quantization}
Quantization is inherently a lossy transformation aimed at reducing model complexity. \format takes an orthogonal approach to model storage by taking the quantized and unquantized models as input, and subsequently performs \textit{lossless} compression to store them efficiently into a unified format. While we focus on storing a pair of models at two specific precisions (e.g. FP16/BF16, INT8) in this paper for brevity, our approach can be generalized to other datatypes (e.g., INT4) and to more-than-two model chains, which we discuss in \cref{sec:implementation_sub} and show in \cref{sec:experiments}.

\subsection{Data Compression}
\label{sec:motivation_compression}
Model hosting platforms (e.g., HuggingFace~\cite{huggingface-transformers}) store foundation models in wrapper formats such as Safetensors~\cite{safetensors, safetensors_usage}, ONNX~\cite{onnx}, TensorFlow,
and SavedModel~\cite{tf_savedmodel} that allow transparent storage of additional information such as tensor names and quantization information along with the model weights. However, these formats store weights in an uncompressed fashion.
Another approach orthogonal to quantization that has been explored to reduce model sizes (for storage) is compression. We discuss the pros and cons of various compression techniques applicable to foundation models.


\paragraph{Generic Compression Algorithms}
Standard compressors such as GZip~\cite{gzip}, ZSTD~\cite{zstd}, LZ4~\cite{lz4} can be applied to model weights. These approaches treat (the sequence of) weights as a generic byte stream and are agnostic to specific structural and numerical properties of the model weights. ALP~\cite{ALP} targets general floating point numbers, but only supports 32-bit and 64-bit floats, so their method cannot be directly applied to 16-bit models. Generic methods do not achieve optimal compression ratios on model weights due to their high entropy (e.g., the mantissa bits of floats~\cite{zipnn}) rendering common techniques such as dictionary coding~\cite{sun2004combining} ineffective.


\paragraph{Compression for ML Models}
Recently, some approaches have been proposed for specifically compressing ML models: ZipNN \cite{zipnn} compresses BF16 weights by reordering the 16-bit float into 2 byte streams and compressing each stream separately with Huffman coding. NeuZip~\cite{neuzip} uses lossy compression to speed up inference by quantizing mantissa bits, and applying lossless compression to exponent bits with an entropy coder to speed up training. Huf-LLM~\cite{huf-llm} uses hardware-aware Huffman compression, breaking the 16-bit value into non-standard bit-level patterns and separately compressing streams for fast inference. MLWeaving stores models in bit-major order, and utilizes FPGA hardware acceleration to enable models to be loaded at arbitrary floating point precisions.

\paragraph{\format (\textbf{ours}): Joint Compression}
Unlike existing compression methods, \format targets the joint storage of a pair of models, and achieves higher overall compression ratios for the model pair versus compressing them separately via delta storage-like techniques~\cite{ilkhechi2020deepsqueeze} for storing the conditional information (\cref{sec:experiments}). Additionally, \format runs purely on CPU, and does not depend on the availability of specific architectures (e.g., systolic arrays, TPUs/NPUs, FPGAs) required by some of the aforementioned methods.

\section{\format Overview}
\label{sec:system}
\begin{figure}[t]
\begin{subfigure}{\columnwidth}
\centering
\begin{tikzpicture}[>={LaTeX[width=1mm,length=1mm]},->]
\tikzset{cbox/.style={
	minimum height=10mm,draw=black,fill=white,ultra thick,align=center,
	minimum width=14mm,anchor=north
}}
\tikzset{dbox/.style={
	minimum height=6mm,minimum width=10mm,draw=black,fill=white,thick,
    font=\scriptsize, align=center,anchor=north,inner sep=0.5mm
}}

\tikzset{
mylabel/.style={
    font=\footnotesize\sffamily\bfseries,
    align=center,
},
mylabel2/.style={
    font=\footnotesize\sffamily,
    align=center,
},
mycomponent/.style={
    semithick, rounded corners=0.5mm,
}
}

\node(inputs) [draw=black, fill = white, mycomponent,align=center,font=\footnotesize,anchor=east, minimum width = 13mm, minimum height = 23.5mm] at (0, 0) {};
\node [anchor=north, align=center, mylabel] at ($(inputs.north)$) {Input};


\node(highprecinput) [draw=black, fill = white, mylabel2,align=center,font=\footnotesize,anchor=north, minimum width = 11mm, minimum height = 2.5mm] at ($(inputs.north) + (0, -0.4)$) {High-\\[-0.2em] precision\\[-0.2em] Model};
\node(lowprecinput) [draw=black, fill = white, mylabel2,align=center,font=\footnotesize,anchor=north, minimum width = 11mm, minimum height = 5mm] at ($(highprecinput.south) + (0, -0.15)$) {Low-\\[-0.2em] precision\\[-0.2em] Model};

\node(encoder) [draw=black, fill = white, mycomponent,align=center,font=\footnotesize\sffamily\bfseries,anchor=west, minimum width = 10mm, minimum height = 10mm] at ($(inputs.east) + (0.4, 0)$) {Encoder\\[-0.2em](\cref{sec:compression_encoding})};

\node(qstore) [draw=black, fill = white, mycomponent,align=center,font=\footnotesize,anchor=west, minimum width = 18mm, minimum height = 23.5mm] at ($(encoder.east) + (0.4, 0)$) {};
\node [anchor=north, align=center, mylabel] at ($(qstore.north) + (0, 0.025)$) {Our unified\\[-0.2em] compression\\[-0.2em] format (\format)\\[-0.2em]};

\node(conditionals) [draw=black, fill = white, mylabel2,align=center,font=\footnotesize,anchor=south, minimum width = 16mm, minimum height = 2.5mm] at ($(qstore.south) + (0, 0.1)$) {conditional\\[-0.2em] weights};
\node(lowprec) [draw=black, inner xsep=0mm, fill = white, mylabel2,align=center,font=\footnotesize,anchor=south, minimum width = 16mm, minimum height = 5mm] at ($(conditionals.north) + (0, 0.1)$) {low-precision\\[-0.2em] model weights};

\node(decoder) [draw=black, fill = white, mycomponent,align=center,font=\footnotesize\sffamily\bfseries,anchor=west, minimum width = 10mm, minimum height = 10mm] at ($(qstore.east) + (0.4, 0)$) {Decoder\\[-0.2em](\cref{sec:compression_decoding})};

\node(outputs) [draw=black, densely dashed, fill = white, mycomponent,align=center,font=\footnotesize,anchor=west, minimum width = 13mm, minimum height = 23.5mm] at ($(decoder.east) + (0.4, 0)$) {};
\node [anchor=north, align=center, mylabel] at ($(outputs.north)$) {Outputs};


\node(highprecoutput) [draw=black, fill = white, mylabel2,align=center,font=\footnotesize,anchor=north, minimum width = 11mm, minimum height = 2.5mm] at ($(outputs.north) + (0, -0.4)$) {High-\\[-0.2em] precision\\[-0.2em] Model};
\node(lowprecoutput) [draw=black, fill = white, mylabel2,align=center,font=\footnotesize,anchor=north, minimum width = 11mm, minimum height = 5mm] at ($(highprecoutput.south) + (0, -0.15)$) {Low-\\[-0.2em] precision\\[-0.2em] Model};

\draw[->, thick] 
($(decoder.east) + (0, -0.3)$) --
($(lowprecoutput.west) + (0, -0.1)$); 
\draw[->, thick] 
($(lowprecoutput.west) + (0, 0.1)$) --
($(decoder.east) + (0, -0.1)$); 
\draw[->, thick] 
($(decoder.east) + (0, 0.1)$) --
($(highprecoutput.west) + (0, 0)$); 

\draw[->, thick] 
($(highprecinput.east) + (0, 0)$) --
($(encoder.west) + (0, 0.1)$); 
\draw[->, thick] 
($(lowprecinput.east) + (0, 0)$) --
($(encoder.west) + (0, -0.1)$); 
\draw[->, thick] 
($(encoder.east) + (0, 0)$) --
($(qstore.west) + (0, -0)$); 
\draw[->, thick] 
($(qstore.east) + (0, 0)$) --
($(decoder.west) + (0, -0)$); 

\end{tikzpicture}
\label{fig:internal_architecture}
\end{subfigure}
\caption{\format pipeline. A high and low-precision model pair is encoded into a unified, storage-friendly format (\format), from which both models can be efficiently retrieved.}
\label{fig:system_overview}
\end{figure}

This section presents the \format pipeline. \format is a format that efficiently stores a pair of high and low-precision models: first, the model pair is compressed using an encoder into the unified \format format. Then, a decoder is applied onto the \format files to losslessly retrieve the high or low-precision model (or both).




\paragraph{QStore Input}
\format's encoding takes in the weights of the high and low-precision model versions ($w$ and $Q(w)$, respectively) as input. \format does not impose restrictions on the input format; our approach can work within any format implementation as long as it stores tensors separately (e.g., safetensors~\cite{safetensors}, PyTorch pickle objects~\cite{pickle}, TensorFlow SavedModel~\cite{tf_savedmodel}, etc. are acceptable).

\paragraph{Encoding}
\format's utilizes an encoder to encode model weights: the encoder first compresses the weights of the low-precision model, then compresses the conditional information present only in the high-precision model (i.e., `extra information', \cref{sec:intro}) (\cref{sec:compression_encoding}).



\paragraph{Format}
The unified \format format, generated by encoding the input model pair, consists of two files: the compressed low-precision weights and the compressed conditional information (\cref{sec:system_format}).


\paragraph{Decoding}
\format's utilizes a decoder to act on the two files within \format to reconstruct either the low or high-precision model (or both): If the user requests the low-precision model, the decoder is invoked on the compressed quantized model weights to reconstruct it. If (additionally) the high-precision model is requested, the decoder is invoked on the newly decompressed low-precision model weights and the compressed conditional information (\cref{sec:compression_decoding}).

\section{\format: Unified Format}
\label{sec:compression}
This section details the \format format and its encoding and decoding algorithms. We describe our intuition to encode conditional information in \cref{sec:compression_preliminary}, the encoding of a model pair into the \format format in \cref{sec:compression_encoding}, the \format format itself in \cref{sec:system_format}, and decoding to obtain the original high or low-precision weights (or both) in \cref{sec:compression_decoding}.

\subsection{Key Intuition}
\label{sec:compression_preliminary} 



This section describes our intuition for compressing conditional information present in the high-precision model but not in the low-precision model. Without loss of generality, we will be describing \format's operations with a FP16/BF16 and INT8 model pair.


\paragraph{Conditional Information} Given a high and low-precision model pair, it is possible to derive the low-precision model from the high-precision model (e.g., via quantization). Hence, all information present in the low-precision model is contained within the high precision model. Given the weights of the high-precision model $W$ and a quantization function $Q$ that maps it to the corresponding quantized weights, we can model the information in the model pair:

\begin{equation}
    H(W) = H(Q(W)) + H(W | Q(W))
    \label{eq:info}
\end{equation}

\format aims to find an efficient bit-level representation corresponding to $H(Q(W)) + H(W | Q(W))$ in \cref{eq:info}. Notably, the representation of the conditional data $W | Q(W)$ must be \textit{lossless} regardless of the quantization function $Q$ used, which \format will not know in advance (i.e., prior to compression). In particular, given floating point $W$ and quantized $Q(W)$, the key challenge is in finding overlapping bit-level patterns in dynamic-precision floating point data that is informed by the corresponding quantized data, which the remainder of this section will aim to address.

\begin{figure}[t]
\usetikzlibrary{patterns}
\begin{subfigure}[b]{\linewidth}
\centering
\begin{tikzpicture}

\pgfplotstableread[col sep=comma,]{
name
\criu
\dumpsession
\system \textbf{(Ours)}
}\datatable

\begin{axis}[
    ybar,
    clip=false,
    xlabel style={yshift =-0.5ex},
    width=90mm,
    height=28mm,
    bar width=5mm,
    ymin=0,
    ymax=12,
    ylabel style={yshift = -3ex,align=center},
    axis y line*=none,
    axis x line*=none,
    ytick={0, 2, 4, 6, 8, 10, 12},
    yticklabels={0, 2, 4, 6, 8, 10, 12},
    xtick={0.5, 1.45, 2.5, 3.55, 4.5},
    xtick style ={draw=none},
    xticklabels={No\\ Grouping, Applied Quant.\\ Function, Post-Quant.\\ Weight, Both (\system), Random\\ Grouping},
    x tick label style={yshift = 1ex},
    xmin=0,
    xmax = 5,
    ymajorgrids,
    tick label style={font=\footnotesize, align=center},
    legend style={
        font=\footnotesize,
        /tikz/every even column/.append style={column sep=0.5cm},
        legend columns = 4,
        at={(-0.15, 1.1)}, anchor=south west
    },
    label style={font=\footnotesize},
    ylabel={Weighted Avg.\\ Entropy},
    area legend
    ]

    \addplot[black,fill=UncompressedColor]coordinates {(1.25,10.41)};
    
    \addplot[black,fill=ZipnnColor]coordinates {(1.85, 10.39)};
    
    \addplot[black,fill=OursColor]coordinates{(2.5, 6.52)};
    \addplot[black,fill=GreenColor]coordinates {(3.15,3.51)};
    \addplot[black,fill=UncompressedColor]coordinates {(3.75,6.87)};

\end{axis}
    
    
\end{tikzpicture}
\end{subfigure}
\caption{Weighted entropy of different grouping strategies on the Llama 3.1 8B Instruct model's 16-bit weights. \format's combined grouping achieves high entropy reduction (hence compression ratio) versus alternative grouping strategies.}
\label{fig:entropy-grouping}
\end{figure}



\paragraph{Grouping by Quantized Weight}
Most common recent quantization schemes use a combination of scaling (e.g., normalizing weights into a range) and rounding (\cref{sec:motivation_quantization}). Given such quantization schemes, we observe that two floats that quantize to the same value (with the same quantization function, described shortly) can be expected to have more overlapping bits compared to two randomly selected floats, such as those that quantize to different values (\cref{fig:entropy-grouping}).
Higher bit-level overlap between floats is directly correlated with compressibility (e.g., via entropy coding schemes); hence, \format groups the high-precision (floats) weights by quantized value during encoding.

\paragraph{Grouping by Quantization Function} Recent popular quantization schemes apply multiple block-wise independent quantization functions to a single tensor (\cref{sec:motivation_quantization}). For example, LLM.int8()~\cite{llmint8dettmers} uses a different scaling factor to quantize each row (e.g., $Q_{row=i}(w_i) = round(\frac{128w_i}{s_i})$, where $s_i$ is row $i$'s scaling factor).
The quantization function is often chosen w.r.t. the weights; a common choice is the magnitude-based $s_i\!=\! abs(max(w_i))$~\cite{llmint8dettmers, awq}. 
Hence, the conditional information of a group of floating point weights w.r.t. their quantized integer weights $H(W|Q(W))$ changes with $Q(W)$. While grouping floats by applied quantization function alone achieves negligible entropy reduction (as intra-group float distributions are still largely random), we observe that a combined grouping of the applied quantization function and quantized weight value achieves significant compression benefits (e.g., versus grouping only by one criteria, or randomly grouping with the same group count, \cref{fig:entropy-grouping}).

\subsection{Encoding to \format}
\label{sec:compression_encoding}
This section describes how a high and low-precision model pair is encoded into the \format format. As described in \cref{sec:system}, \format's encoder compresses the low-precision model and the high-precision model's conditional information w.r.t. low-precision model (\cref{sec:compression_preliminary}). 


\paragraph{Encoding Quantized Weights}
\format's encoder utilizes an entropy coding scheme to compress the (quantized) weights of the low-precision model $Q(w)$. It follows zstd's approach~\cite{zstd} to divide $Q(w)$ into sequential, fixed-size chunks, on which per-chunk Huffman compression is applied for up to 12\% size reduction (\cref{sec:experiment_space}). 

\paragraph{Encoding Conditional Information}
\format's encoder computes conditional information using weights of both the high and low-precision model ($w$ and $Q(w)$) as input. Following intuition described in \cref{sec:compression_preliminary}, the high-precision model weights $w$ are first grouped according to applied quantization function (e.g., for LLM.int8()~\cite{llmint8dettmers}, each group will consist of tensors with the same applied scale value). Then, weights in each group are further divided into subgroups of weights quantizing to the same value. For example, in \ref{fig:qstore}, rows $w_1$, $w_3$, and $w_2$ are quantized with distinct scale values ($32$ and $16$), hence their weights are placed into groups 1 ($s_1 = s_3 = 32$) and 2 ($s_2 = 16$). In group 1, $w_{11}, w_{13}, w_{32}$, and $w_{33}$ quantize to the same value (yellow) and are placed in one subgroup; $w_{12}$ and $w_{32}$ quantize to another value (blue) and are placed in another subgroup.

\paragraph{Per-subgroup compression}
Similar to how we compress the low-precision quantized weights, \format's conditional encoder then compresses conditional information using Huffman compression on a per-subgroup basis. If a chunk is not compressible enough (e.g., due to high entropy, or very few unique values in a subgroup), \format skips encoding and stores that chunk uncompressed.

\paragraph{Remark}
The combined size of \format's compressed quantized weights and conditional information is much lower than the original uncompressed size of both models; in fact, \format's size is close to \textit{only} compressing the high-precision model (e.g., via ZipNN, \cref{sec:experiment_space}); however, \format additionally allows the low-precision model to be directly retrieved without requiring in-memory quantization (\cref{sec:exp_online_quantization}).

\input{plots/fig_qstore}

\subsection{\format Format}
\label{sec:system_format}
This section describes how \format stores an encoded high and low-precision model pair. Each \format model pair consists of two files---the compressed quantized weights and conditional information.


\paragraph{Compressed Quantized Weights} 
\format stores compressed low-precision model weights alongside a header---chunk count, tensor dimensions, and per-chunk metadata of (1) whether compression was applied and (2) compressed and uncompressed chunk sizes. 

\paragraph{Compressed Conditional Information} 
\format stores conditional information following group (i.e., quantization function), then subgroup (i.e., post-quantization value) order. It maintains a header, which stores (1) group-to-position mappings in the original model (e.g., row number), and within each group, (2) the aforementioned per-subgroup data. Notably, despite \format also reordering the weights in each group based on subgroups, it does not store per-subgroup (row) weight position mappings: this is because the quantized weights already contain the information, e.g., $w_{13}$ assigned to group 1, subgroup 1 in \cref{fig:qstore} can be inferred to be row $w_1$'s $3^rd$ element based on the corresponding quantized weights in $Q_1(w_1)$.

\subsection{Decoding from \format}
\label{sec:compression_decoding}
This section covers how a model pair stored with \format can be losslessly decoded to retrieve the high and/or low-precision models.


\paragraph{Retrieving the Low-Precision Model}
The model's quantized weights are encoded to \format with per-chunk Huffman compression into a file (\cref{sec:compression_encoding}). Hence, directly loading the compressed quantized weights from \format, and applying per-chunk huffman decompression allows the low-precision model to be retrieved losslessly.

\paragraph{Retrieving the High-Precision Model}
As \format stores the encoded conditional information for the high-precision model w.r.t. the low-precision model, it requires the low-precision model to be retrieved first. Then, \format's decoder first decompresses the conditional information, which is applied onto the low-precision model weights to retrieve correct per-group weight ordering (\cref{sec:system_format}. Finally, \format uses the stored group-to-row mappings to losslessly reconstruct the high-precision model's weight tensor.



\paragraph{Remark}
\format's loading of the high or low-precision model is faster than loading the respective model uncompressed, and comparable to loading the respective model (separately) compressed with an off-the-shelf tool (e.g., ZipNN, \cref{sec:experiment_decode}). However, as \format jointly stores the model pair, \format achieves significant time savings for loading the low-precision model versus the common practice of loading the unquantized model, then quantizing it in memory (\cref{sec:exp_online_quantization}).

\section{Implementation \& Discussion}
\label{sec:implementation_sub}
\paragraph{Choice of Encoding Scheme}
Our implementation of \format uses the FiniteStateEntropy library's near-state-of-the-art Huffman encoding Huff0~\cite{fse}. However, other entropy-based encoding schemes can be used instead, such as the FiniteStateEntropy coder from the same library or non-Huffman methods. (e.g., arithmetic coding~\cite{arithmetic_coding})


\paragraph{Efficient Decode Pipelining}
We pipeline \format's per-tensor decoding for model loading (\cref{sec:compression_decoding}), where one tensor's decompression overlaps with the next tensor's read. However, other parallelization strategies can be used in its place~\cite{losslessdecomp, comp_interleave}, such as completely parallelizing tensor read and decompression, which may bring larger benefits on specific hardware (e.g., local SSD~\cite{cao2017performance, shriver1999does}).


\paragraph{Lazy Model Loading}
As \format's encoding and decoding of model pairs operate independently on each tensor, it can be naturally extended to support lazy loading (e.g., similar to Safetensors~\cite{safetensors}). For lazy loading, we would not apply decode pipelining, and only read and decompress tensors when required; we defer detailed performance optimization and engineering to future work.

\paragraph{Generalizing to Multiple Precisions \& Datatypes}
While the use of model pairs is common, many modern deployments require models in more than two quantization formats (e.g., 4-bit mobile inference, 8-bit cloud inference, and 16-bit training). In such situations, all low precision models (4-bit, 8-bit) are created by quantizing the high precision (16-bit) model. \format can be extended to store more-than-two model chains, only requiring that the same group size is used for the quantization to different precisions (verified in \cref{sec:exp_multichain}): for the prior example, \format would store the INT4 model, the INT8 | INT4 conditional encoding, and the FP16 | INT8 conditional encoding. As mentioned in \cref{sec:intro}, this extension would especially benefit model storage hubs like HuggingFace~\cite{huggingface-transformers} which can store multiple quantized representations of the same model with significantly lower storage cost versus separately storing precisions.

\paragraph{Compatibility with Precisions and Datatypes} 
While we evaluate \system with LLM.int8() and GPT-Q~\cite{gptq} quantization and the FP16/BF16/INT8/INT4 datatypes (\cref{sec:experiments}), \format can be extended to support other datatypes and group-based quantization methods: \format directly applies byte-level entropy coding for storage (\cref{sec:compression_encoding}); only the group-wise weight ordering (in the low-precision model) and conditional information are required to losslessly reconstruct the high-precision model (\cref{sec:compression_decoding}), both of which are datatype-agnostic.

\section{Evaluation} 
\label{sec:experiments}
In this section, we empirically study the effectiveness of \format's quantization-aware model storage. We make the following claims:

\begin{table}[t]
\midsepremove
\caption{Summary of models used for evaluation.}
\footnotesize
\begin{tabular}{lrrr}
\toprule
\textbf{Model}  & \textbf{Params.} & \textbf{Model Pair Size} & \textbf{Modality}\\
 \midrule
Qwen 2 Audio~\cite{qwen-audio}  & 7B & 19.9 GB & Audio-Text \\
Mistral v0.3~\cite{mistral}  & 7B & 19.5 GB & Text \\
Llama 3.1~\cite{llama3}  & 8B & 19.5 GB & Text \\
\midrule
Gemma 3~\cite{gemma3}  & 27B & 72.7 GB & Image-Text \\
Qwen 2.5 VL~\cite{Qwen2.5-VL}  & 32B & 87.7 GB & Video-Image-Text \\
Deepseek Coder~\cite{deepseek-coder}  & 33B & 91.9 GB & Text (Coding) \\
\bottomrule
\end{tabular}
\midsepdefault
\label{tbl:workload}
\end{table}

\noindent
\begin{enumerate}
    \item \textbf{Effective Compression:} \format achieves up to $2.2\times$ compression ratio for storing a high and low-precision model pair---up to $1.6\times$ better than the next best method. (\cref{sec:experiment_space})

    \item \textbf{Fast Retrieval:} A model pair stored with \format can be loaded up to $1.8\times$ faster versus alternative formats (\cref{sec:experiment_decode}).
        \item \textbf{Fast Storage:} A model pair can be stored with \format up to $2.8\times$ faster than uncompressed storage, and $1.7\times$ faster versus alternative storage/compression methods (\cref{sec:experiment-encode}).    
    
    
\end{enumerate}
\noindent
\begin{table*}[t]

\caption{\format's storage cost (GBs) for storing a high and low precision model pair versus baselines. \format achieves up to $2.2\times$ and $1.6\times$ space savings versus storing the models uncompressed (\safetensors) and the next best alternative (\zipnn + \zstd).}
\footnotesize
\addtolength{\tabcolsep}{-1.3pt} 
\midsepremove
\begin{tabular}{|l|r|r|r|r|r|r|r|r|r|r|r|r|r|r|r|r|r|r|r|r|}
\toprule
 \multicolumn{1}{|r|}{\textbf{Method}} & \multicolumn{3}{c|}{\safetensors}& \multicolumn{3}{c|}{\lzfour} & \multicolumn{3}{c|}{\zstd}& \multicolumn{3}{c|}{\zipnn (high prec) + \zstd (low prec)} & \multicolumn{3}{c|}{\system \textbf{(Ours)}} \\
  \midrule
\textbf{Model} & \multicolumn{1}{|c|}{Low} & \multicolumn{1}{|c|}{High} & \multicolumn{1}{|c|}{\cellcolor{gray!20}\textbf{Total}} & \multicolumn{1}{|c|}{Low} & \multicolumn{1}{|c|}{High} & \multicolumn{1}{|c|}{\cellcolor{gray!20}\textbf{Total}} & \multicolumn{1}{|c|}{Low} & \multicolumn{1}{|c|}{High} & \multicolumn{1}{|c|}{\cellcolor{gray!20}\textbf{Total}}& \multicolumn{1}{|c|}{Low (\zstd)} & \multicolumn{1}{|c|}{High (\zipnn)} & \multicolumn{1}{|c|}{\cellcolor{gray!20}\textbf{Total}} & \multicolumn{1}{|c|}{Low} & \multicolumn{1}{|c|}{\textbf{High$|$Low}} & \multicolumn{1}{|c|}{\cellcolor{gray!20}\textbf{Total}} \\
 \midrule
Qwen 2 Audio 7B

 &6.622 & 13.244 & \cellcolor{gray!20} 19.866 &  6.647  &13.296&  \cellcolor{gray!20}19.943  &  5.891 & 10.303& \cellcolor{gray!20}16.194  & 5.891 & 8.780 & \cellcolor{gray!20}14.671 & 5.893 & 3.572 & \cellcolor{gray!20}\textbf{9.465}\\
\midrule
Mistral v0.3 8B & 6.500 & 13.000 & \cellcolor{gray!20}19.500 &6.524 &13.051  &  \cellcolor{gray!20}19.575& 5.734 & 10.124 & \cellcolor{gray!20}15.858& 5.734 & 8.629 &  \cellcolor{gray!20}14.363& 5.734 & 3.379 &\cellcolor{gray!20} \textbf{9.113}  \\
\midrule
Llama 3.1 8B  & 6.500 & 13.000 & \cellcolor{gray!20} 19.500 & 6.523& 13.051& \cellcolor{gray!20}19.574 & 5.746 & 10.083 & \cellcolor{gray!20}15.829 & 5.746& 8.629 & \cellcolor{gray!20}14.375& 5.746 & 3.295 & \cellcolor{gray!20} \textbf{9.041}\\
\midrule
Gemma 3 27B
  & 24.223 &  48.446 &   \cellcolor{gray!20}72.669 & 24.317 & 48.636 &\cellcolor{gray!20} 72.954 & 21.230 & 37.450 &\cellcolor{gray!20}58.680 & 21.230 & 32.153 &\cellcolor{gray!20}53.383 & 21.230 & 11.848 & 
 \cellcolor{gray!20} \textbf{33.078} \\
\midrule
Qwen 2.5 VL 32B
 &  29.248 & 58.496 & \cellcolor{gray!20}87.744 & 29.164 & 58.718 & \cellcolor{gray!20}87.882 & 25.226 & 44.869 &\cellcolor{gray!20}70.095 & 25.226 & 39.013 &\cellcolor{gray!20}64.239 & 25.295 & 13.940 &\cellcolor{gray!20}\textbf{39.235}  \\
\midrule
Deepseek Coder 33B  & 30.621 & 61.243 &\cellcolor{gray!20} 91.864& 30.697& 61.483 & \cellcolor{gray!20} 92.180 &27.030  & 47.959 & \cellcolor{gray!20}74.989 &27.030 & 40.572 & \cellcolor{gray!20} 67.602& 27.041 & 14.548 & \cellcolor{gray!20} \textbf{41.589} \\
\bottomrule
\end{tabular}
\midsepdefault
\addtolength{\tabcolsep}{1.3pt}
\label{tbl:exp_size_both}
\end{table*}

\textbf{Deeper Performance analysis of \format (Ours)}
\begin{enumerate}
    \item \textbf{Effective Under Constrained Bandwidth:} \format enables up to $2.2\times$ faster model loading versus loading uncompressed models under I/O-constrained scenarios (\cref{sec:exp_bandwidth}).

    \item\textbf{Comparison with Online Quantization:} The low-precision model can be loaded from \format up to $2.5\times$ faster versus loading and quantizing a high-precision model (\cref{sec:exp_online_quantization}).

    
    \item \textbf{Effective Compression of Multi-Level Model Chains:} 
    \format achieves up to $2.46\times$ compression ratio when storing model chains with more than two precision levels; this is up to $1.78\times$ better than the next best alternative. (\cref{sec:exp_multichain})
    
\end{enumerate}

\subsection{Experimental Setup}
\label{sec:experiment_setup}

\paragraph{Dataset (\cref{tbl:workload})}
We select 6 popular foundation models for evaluation, which we further divide 
into 3 `small' (<20B parameters) and 3 `large' ($\geq$20B parameters) models. For each model, we create a high and low-precision model pair consisting of the (1) original BF16 model and (2) quantized INT8 model (via LLM.int8()~\cite{llmint8dettmers}, unless otherwise stated, e.g., in \cref{sec:exp_online_quantization} and \cref{sec:exp_multichain}) weights.


\paragraph{Methods}
We evaluate \format against existing tools and methods capable of storing the high and low-precision model pairs:
\begin{itemize}
    \item \textbf{\safetensors~\cite{safetensors}:} The default uncompressed model storage format~\cite{safetensors_load_function, safetensors_without_lazy_loading} of HuggingFace's transformers library~\cite{huggingface-transformers}.
    \item \textbf{\lzfour~\cite{lz4}:} We use the default compression level of 1.
    \item \textbf{\zstd~\cite{zstd}:} We use a compression level of 2.
    \item \textbf{\zipnn~\cite{zipnn}}: A Huffman-based compression algorithm that targets compression of 16-bit model weights. Since it cannot compress 8-bit weights, in order to compare the storage cost of both precisions, we use \zipnn for high precision and the best alternative baseline (\zstd) for low precision.
\end{itemize}
We implement all methods to sequentially process tensors to and from a single file for model saving and loading. Tensor read/write and (de)compression are pipelined to overlap I/O and compute (\cref{sec:implementation_sub}).


\paragraph{Environment}
We use an Azure Standard E80is (Intel(R) Xeon Platinum 8272CL, 64-bit, little-endian) VM instance with 504GB RAM. We read and write (compressed) model data to and from local SSD for all methods. 
The disk read and write speeds are 1.5 GB/s and 256.2 MB/s, respectively,\footnote{Measured with $dd$ with $1MB$ block size, reading $1024$ blocks from a model file.} with read latency of 7.49ms.\footnote{Measured with \texttt{iostat -x}.}


\paragraph{Time Measurements}
We measure (1) \textit{save time} to compress and store a model pair onto storage, and (2) \textit{load time} to read and decompress the selected model(s) into memory.
We force full data writing (via \texttt{sync}~\cite{sync}) and reading during model saving and loading.
We perform data reading and writing with a single thread and compression/decompression with 48 threads for all methods. 


\paragraph{Reproducibility} Our implementation of \format and experiment scripts can be found in our \href{https://github.com/illinoisdata/qstore}{\textcolor{blue}{Github repository}}.

\subsection{\format Saves Model Storage Cost}
\label{sec:experiment_space}

This section studies \format's model storage cost savings. We store model pairs to disk with each method, and compare the resulting on-disk file sizes of \format versus alternative methods in \cref{tbl:exp_size_both}.


\format's file size is consistently smallest, and is up to $2.2\times$ and $1.6\times$ smaller versus \safetensors (uncompressed) and next best compression method (\zipnn + \zstd), respectively. As hypothesized in \cref{sec:motivation_compression}, \zstd and \lzfour achieve suboptimal compression ratios due to the traditional compression techniques they utilize being ineffective on noisy, high-entropy model tensor data---notably, \lzfour achieves \textit{no} benefits storage-wise.
While \zipnn effectively compresses just the high-precision model into a size smaller (up to $6.3\%$) versus \system's model pair, its specialization for the FP16/BF16 data formats leads to it having to rely on a different, less effective compression algorithm (\zstd) for storing the low-precision model, leading to \system's stored model pair being $1.6\times$ smaller than the pair stored with \zipnn + \zstd.
\system's high compression ratio translates to significant (52\%-55\%) space savings across model sizes: storing the Deepseek Coder's model pair with \format takes only 42GB versus the 92GB of storing the models as is without compression.

\paragraph{Effective Conditional Information Storage} \format's compressed conditional information (\textit{High$|$Low}) only takes up to $39\%$ of the total size, and accordingly contributes only up to $40\%$ of the model pair loading time (\cref{fig:exp_decode_both}, \cref{fig:exp_decode_bf16}) across all 6 models.
This shows \format's conditional encoding's effectiveness in reducing storage and load time redundancies incurred by the typical approach of users storing and using both the high and low-precision models as is (\cref{sec:intro}).

\subsection{\format Saves Model Load Time}
\label{sec:experiment_decode}

We investigate \format's time savings for loading a model pair. We store the model pair using each method, then measure the time taken for loading one or both models from storage into memory. 

We report results for loading both models in \cref{fig:exp_decode_both}. \format saves significant time in cases where simultaneous access to both models (e.g., model cascade and chaining~\cref{sec:intro} or interactive computing~\cite{kishu,kishudemo,elasticnotebook,elasticnotebook-demo}) is required; it loads the model pair up to $2.2\times$ and $1.8\times$ faster than separately loading the two models stored without compression (\safetensors) or with an applicable compression algorithm (\zstd), respectively; one significant cause of faster loading is the \format's model pair size being significantly smaller than that of by separately storing the two models with alternative approaches (\cref{sec:experiment_space}), which saves I/O costs especially under constrained bandwidths (\cref{sec:exp_bandwidth}). 

\paragraph{Comparable High-Precision Model Load Time (\cref{fig:exp_decode_bf16})}
\format loads the high-precision model up to $1.4\times$ faster versus loading uncompressed (\safetensors), and has comparable loading times versus loading it with a specialized method ($\pm 5\%$, \zipnn). 
\system still saves storage space in this case as it jointly stores the low-precision model (when it is needed): While storing the low-precision model is not required for the latter to load the high-precision model (unlike \system), the alternative of not storing the low-precision model can result in high online quantization costs (\cref{sec:exp_online_quantization}).

\newcommand\figwidth{45mm}

\begin{figure}[t]\captionsetup[subfigure]{font=footnotesize}
\pgfplotsset{scaled y ticks=false}
\centering
\begin{subfigure}[b]{0.48\linewidth}
\begin{tikzpicture}

\begin{axis}[
    xtick=data,
    clip=false,
    width=\figwidth,
    height=28mm,
    ymin=0,
    ymax=30,
    axis y line*=none,
    axis x line*=none,
    ytick={0, 10, 20, 30},
    yticklabels={0, 10, 20, 30},
    xlabel=Storage size (GB),
    xlabel style={yshift = 3ex},
    label style={font=\scriptsize},
    ylabel style={yshift=-4.5ex,xshift=-0.5ex, font=\scriptsize},
    xmin = 0,
    xmax = 20,
    xtick = {0, 5, 10, 15, 20},
    xticklabels = {0, 5, 10, 15, 20},
    tick label style={font=\scriptsize},
    x tick label style={yshift=0.5ex},
    legend style={
        at={(-0.3,1.07)},anchor=south west,column sep=2pt,
        row sep = -0.4pt,
        draw=black,fill=white,
        inner ysep=0.1pt,
        /tikz/every even column/.append style={column sep=0.2cm},
        font=\footnotesize,
        font=\scriptsize
    },
    legend cell align={left},
    legend columns=5,
    ylabel={Load time (s)},
    ymajorgrids,
    every axis plot/.append style={thick}
]

\addplot[only marks, mark = *, mark size = 1.5pt, BaseBothColor]
table[x=x,y=y] {
x y
19.866 26.1508
19.500 25.564
19.500 25.540
};
\addlegendentry{\safetensors}
\addplot[only marks, mark = +, mark size = 1.5pt, LzfourColor]
table[x=x,y=y] {
x y
19.943 27.792
19.575 27.998
19.574 27.816
};
\addlegendentry{\lzfour}
\addplot[only marks, mark=x, mark size = 1.5pt, ZstdColor]
table[x=x,y=y] {
x y
16.194 21.857
15.858 22.542
15.829 22.379
};
\addlegendentry{\zstd}
\addplot[only marks, mark= square*,mark size = 1.5pt, ZipnnColor]
table[x=x,y=y] {
x y
14.671  20.406
14.363 20.857
14.375 18.924
};
\addlegendentry{\zipnn (h. prec) + \zstd (l. prec)}
\addplot[only marks, mark=triangle*, mark size = 2.5pt, OursColor]
table[x=x,y=y] {
x y
9.465 12.054
9.113 11.773
9.041 12.024
};
\addlegendentry{\system \textbf(Ours)}

\end{axis}
\end{tikzpicture}
\vspace{-6.5mm}
\caption{Small Models (<20B Params.)}
\end{subfigure}
\begin{subfigure}[b]{0.48\linewidth}
\begin{tikzpicture}

\begin{axis}[
    xtick=data,
    clip=false,
    width=\figwidth,
    height=28mm,
    ymin=0,
    ymax=150,
    axis y line*=none,
    axis x line*=none,
    ytick={0, 50, 100, 150},
    yticklabels={0, 50, 100, 150},
    xlabel=Storage size (GB),
    xlabel style={yshift = 3ex},
    label style={font=\scriptsize},
    ylabel style={yshift=-4.5ex,xshift=-0.5ex, font=\scriptsize},
    xmin = 0,
    xmax = 100,
    xtick = {0, 25, 50, 75, 100},
    xticklabels = {0, 25, 50, 75, 100},
    tick label style={font=\scriptsize},
    x tick label style={yshift=0.5ex},
    legend style={
        at={(0.05,1.17)},anchor=south west,column sep=2pt,
        row sep = -0.4pt,
        draw=black,fill=white,
        inner ysep=0.1pt,
        /tikz/every even column/.append style={column sep=10pt},
        font=\footnotesize
    },
    legend cell align={left},
    legend columns=8,
    ylabel={Load Time (s)},
    ymajorgrids,
    every axis plot/.append style={thick}
]

\addplot[only marks, mark = *, mark size = 1.5pt, BaseBothColor]
table[x=x,y=y] {
x y
72.869 97.505
87.744 114.298
91.864 120.849
};
\addplot[only marks, mark = +, mark size = 1.5pt, LzfourColor]
table[x=x,y=y] {
x y
72.954 104.184
87.882 126.482
92.180 134.328
};
\addplot[only marks, mark=x, mark size = 1.5pt, ZstdColor]
table[x=x,y=y] {
x y
58.860 84.891
70.095 101.149
74.989 109.549
};
\addplot[only marks, mark= square*,mark size = 1.5pt, ZipnnColor]
table[x=x,y=y] {
x y
53.383 77.721
64.239 95.022
67.602 99.775
};
\addplot[only marks, mark=triangle*, mark size = 2.5pt, OursColor]
table[x=x,y=y] {
x y
33.078 47.905
39.235 56.097
41.589 61.357
};

\end{axis}
\end{tikzpicture}
\vspace{-2.5mm}
\caption{Large Models (>20B Params.)}
\end{subfigure}

\caption{Decoding time when we need both high-precision and low-precision models, versus storage costs: \format's loads the model pair up to $2.2\times$ and $1.8\times$ faster versus loading uncompressed models and compression baselines.}
\label{fig:exp_decode_both}
\end{figure}
\begin{figure}[t]\captionsetup[subfigure]{font=footnotesize}
\pgfplotsset{scaled y ticks=false}
\centering
\begin{subfigure}[b]{0.48\linewidth}
\begin{tikzpicture}

\begin{axis}[
    xtick=data,
    clip=false,
    width=\figwidth,
    height=28mm,
    ymin=0,
    ymax=20,
    axis y line*=none,
    axis x line*=none,
    ytick={0, 5, 10, 15, 20},
    yticklabels={0, 5, 10, 15, 20},
    xlabel=Storage size (GB),
    xlabel style={yshift = 3ex},
    label style={font=\scriptsize},
    ylabel style={yshift=-4.5ex,xshift=-0.5ex, font=\scriptsize},
    xmin = 0,
    xmax = 20,
    xtick = {0, 5, 10, 15, 20},
    xticklabels = {0, 5, 10, 15, 20},
    tick label style={font=\scriptsize},
    x tick label style={yshift=0.5ex},
    legend style={
        at={(-0.3,1.07)},anchor=south west,column sep=2pt,
        row sep = -0.4pt,
        draw=black,fill=white,
        inner ysep=0.1pt,
        /tikz/every even column/.append style={column sep=0.2cm},
        font=\footnotesize,
        font=\scriptsize
    },
    legend cell align={left},
    legend columns=5,
    ylabel={Load time (s)},
    ymajorgrids,
    every axis plot/.append style={thick}
]

\addplot[only marks, mark = *, mark size = 1.5pt, BaseBothColor]
table[x=x,y=y] {
x y
19.866  17.3468
19.500 17.0119
19.500 17.2354
};
\addlegendentry{\safetensors}
\addplot[only marks, mark = +, mark size = 1.5pt, LzfourColor]
table[x=x,y=y] {
x y
19.943 18.917
19.575 18.818
19.574 18.561 
};
\addlegendentry{\lzfour}
\addplot[only marks, mark=x, mark size = 1.5pt, ZstdColor]
table[x=x,y=y] {
x y
16.194 13.808
15.858  14.333
15.829 14.193
};
\addlegendentry{\zstd}
\addplot[only marks, mark= square*,mark size = 1.5pt, ZipnnColor]
table[x=x,y=y] {
x y
14.671 11.602
14.363 12.305
14.375 12.214
};
\addlegendentry{\zipnn (h. prec) + \zstd (l. prec)}
\addplot[only marks, mark=triangle*, mark size = 2.5pt, OursColor]
table[x=x,y=y] {
x y
9.465 12.054
9.113 11.773
9.041 12.024
};
\addlegendentry{\system \textbf(Ours)}

\end{axis}
\end{tikzpicture}
\vspace{-6.5mm}
\caption{Small Models (<20B Params.)}
\end{subfigure}
\begin{subfigure}[b]{0.48\linewidth}
\begin{tikzpicture}

\begin{axis}[
    xtick=data,
    clip=false,
    width=\figwidth,
    height=28mm,
    ymin=0,
    ymax=100,
    axis y line*=none,
    axis x line*=none,
    ytick={0, 25, 50, 75, 100},
    yticklabels={0, 25, 50, 75, 100},
    xlabel=Storage size (GB),
    xlabel style={yshift = 3ex},
    label style={font=\scriptsize},
    ylabel style={yshift=-4.5ex,xshift=-0.5ex, font=\scriptsize},
    xmin = 0,
    xmax = 100,
    xtick = {0, 25, 50, 75, 100},
    xticklabels = {0, 25, 50, 75, 100},
    tick label style={font=\scriptsize},
    x tick label style={yshift=0.5ex},
    legend style={
        at={(0.05,1.17)},anchor=south west,column sep=2pt,
        row sep = -0.4pt,
        draw=black,fill=white,
        inner ysep=0.1pt,
        /tikz/every even column/.append style={column sep=10pt},
        font=\footnotesize
    },
    legend cell align={left},
    legend columns=8,
    ylabel={Load Time (s)},
    ymajorgrids,
    every axis plot/.append style={thick}
]

\addplot[only marks, mark = *, mark size = 1.5pt, BaseBothColor]
table[x=x,y=y] {
x y
72.869 65.5253
87.744 76.153
91.864 80.572
};
\addplot[only marks, mark = +, mark size = 1.5pt, LzfourColor]
table[x=x,y=y] {
x y
72.954 70
87.882 84.276
92.180 89.706
};
\addplot[only marks, mark=x, mark size = 1.5pt, ZstdColor]
table[x=x,y=y] {
x y
58.860 54.726
70.095 65.366
74.989 70.452
};
\addplot[only marks, mark= square*,mark size = 1.5pt, ZipnnColor]
table[x=x,y=y] {
x y
53.383 45.741
64.239 56.877
67.602 59.498
};
\addplot[only marks, mark=triangle*, mark size = 2.5pt, OursColor]
table[x=x,y=y] {
x y
33.078 47.905
39.235 56.097
41.589 61.357
};

\end{axis}
\end{tikzpicture}
\vspace{-2.5mm}
\caption{Large Models (>20B Params.)}
\end{subfigure}

\caption{Decoding time when we only need high-precision models, versus storage costs: \system loads the model up to $1.4\times$ faster versus loading uncompressed, comparable ($\pm 5\%$) to loading with a specialized compression algorithm (\zipnn).}
\label{fig:exp_decode_bf16}
\end{figure}

\begin{figure}[t]\captionsetup[subfigure]{font=footnotesize}
\usetikzlibrary{patterns}
\begin{subfigure}[b]{0.48\linewidth}
\centering
\begin{tikzpicture}

\pgfplotstableread[col sep=comma,]{
name
Qwen 7B
Mistral 8B
Llama 8B
}\datatable

\begin{axis}[
    ybar,
    clip=false,
    xtick={1, 2, 3},
    xticklabels from table={\datatable}{name},
                 x tick label style={anchor=center, yshift = 0ex, font=\scriptsize},
    xtick style ={draw=none},
    xlabel style={yshift = 2.5ex, font=\footnotesize},
    ylabel style={yshift = -4.5ex, font=\scriptsize},
    width=45mm,
    height=28mm,
    bar width=1.0mm,
    ymin=0,
    ymax=150,
    axis y line*=none,
    axis x line*=none,
    ytick={0, 50, 100, 150},
    yticklabels={0, 50, 100, 150},
    xmin=0.5,
    xmax = 3.5,
    ymajorgrids,
    tick label style={font=\footnotesize},
    legend style={
        font=\footnotesize,
        /tikz/every even column/.append style={column sep=0.2cm},
        legend columns = 6,
        at={(-0.3,1.1)},
                inner ysep=0.1pt,
        anchor=south west,
        legend image post style={xscale=0.3},
        font=\scriptsize
    },
    ylabel={Encode Time (s)},
    area legend
    ]
                 
    \addplot [black,fill=BaseBothColor,x tick label style={xshift=-0.3cm}, postaction={
        pattern=north west lines
    }] table[x=Method,y=baseboth] {data/exp_encode_small.txt};
    \addlegendentry[]{\safetensors};

    
    \addplot [black,fill=LzfourColor,x tick label style={xshift=-0.3cm}, postaction={
        pattern=grid
    }] table[x=Method,y=lzfour] {data/exp_encode_small.txt};
    \addlegendentry[]{\lzfour};
    
    \addplot
    [black,fill=ZstdColor,x tick label style={xshift=-0.3cm}, postaction={
        pattern=crosshatch dots
    }] table[x=Method,y=zstd] {data/exp_encode_small.txt};
    \addlegendentry[]{\zstd};

    \addplot [black,fill=ZipnnColor,x tick label style={xshift=-0.3cm}, postaction={
        pattern=bricks
    }] table[x=Method,y=zipnn] {data/exp_encode_small.txt};
    \addlegendentry[]{\zipnn (h. prec) + \zstd (l. prec)};
    
    \addplot [black,fill=OursColor,x tick label style={xshift=-0.3cm}, postaction={
        pattern=crosshatch
    }] table[x=Method,y=ours] {data/exp_encode_small.txt};
    \addlegendentry[]{\format \textbf{(Ours)}};

\end{axis}
    
\end{tikzpicture}
\vspace{-7mm}
\caption{Small Models (<20B Params.)}
\label{fig:exp_encode_small}
\end{subfigure}
\begin{subfigure}[b]{0.48\linewidth}
\centering
\begin{tikzpicture}

\pgfplotstableread[col sep=comma,]{
name
Gemma
Qwen 32B
Deepseek
}\datatable

\begin{axis}[
    ybar,
    clip=false,
    xtick={1, 2, 3},
    xticklabels from table={\datatable}{name},
                 x tick label style={anchor=center, yshift = 0ex, font=\scriptsize},
    xtick style ={draw=none},
    xlabel style={yshift = 2.5ex, font=\footnotesize},
    ylabel style={yshift = -4.5ex, font=\scriptsize},
    width=45mm,
    height=28mm,
    bar width=1.0mm,
    ymin=0,
    ymax=630,
    axis y line*=none,
    axis x line*=none,
    ytick={0, 200, 400, 600},
    yticklabels={0, 200, 400, 600},
    xmin=0.5,
    xmax = 3.5,
    ymajorgrids,
    tick label style={font=\footnotesize},
    legend style={
        font=\footnotesize,
        /tikz/every even column/.append style={column sep=0.5cm},
        legend columns = 6,
        at={(0,1.1)},
        anchor=south west,
    },
    ylabel={Encode Time (s)},
    area legend
    ]
                 
    \addplot [black,fill=BaseBothColor,x tick label style={xshift=-0.3cm}, postaction={
        pattern=north west lines
    }] table[x=Method,y=baseboth] {data/exp_encode_large.txt};

    
    \addplot [black,fill=LzfourColor,x tick label style={xshift=-0.3cm}, postaction={
        pattern=grid
    }] table[x=Method,y=lzfour] {data/exp_encode_large.txt};
    
    \addplot
    [black,fill=ZstdColor,x tick label style={xshift=-0.3cm}, postaction={
        pattern=crosshatch dots
    }] table[x=Method,y=zstd] {data/exp_encode_large.txt};

    \addplot [black,fill=ZipnnColor,x tick label style={xshift=-0.3cm}, postaction={
        pattern=bricks
    }] table[x=Method,y=zipnn] {data/exp_encode_large.txt};
    
    \addplot [black,fill=OursColor,x tick label style={xshift=-0.3cm}, postaction={
        pattern=crosshatch
    }] table[x=Method,y=ours] {data/exp_encode_large.txt};

\end{axis}
    
\end{tikzpicture}
\vspace{-3mm}
\caption{Large Models (>20B Params.)}
\label{fig:exp_encode_large}
\end{subfigure}

\caption{\format's encoding time for saving a model pair versus baselines. \format saves the models $2.8\times$ faster versus uncompressed storage, and is up to $1.7\times$ faster than storing the models with an applicable compression algorithm.
}
\label{fig:exp_encode}
\end{figure}
\begin{table}[t]

\caption{Average bits per weight to store each model pair.}
\midsepremove
\footnotesize
\begin{tabular}{lccc}
\toprule
\textbf{Model}  & \textbf{\safetensors} & \textbf{\zstd} & \textbf{\format (Ours)}\\
 \midrule
Qwen 2 Audio~\cite{qwen-audio}  & 24 & 19.564 & \textbf{11.434} \\
Mistral v0.3~\cite{mistral}  & 24 & 19.518 & \textbf{11.216} \\
Llama 3.1~\cite{llama3}  & 24 & 19.482 & \textbf{11.127} \\
Gemma 3~\cite{gemma3}  & 24 & 19.379 & \textbf{10.925} \\
Qwen 2.5 VL~\cite{Qwen2.5-VL}  & 24 & 19.173 & \textbf{10.732} \\
Deepseek Coder~\cite{deepseek-coder}  & 24 & 19.591 & \textbf{10.865} \\
\bottomrule
\end{tabular}
\midsepdefault
\label{tbl:bits}
\end{table}



\subsection{\format Enables Faster Model Storage}
\label{sec:experiment-encode}
This section investigates \format's time for storing model pairs. We measure the time taken for storing a model pair from memory into storage with the \format format versus alternative methods. 

We report results in \cref{fig:exp_encode}. \format's model pair storing time is up to $1.7\times$ and $2.8\times$ faster compared to the next best compression scheme and non-compression method, respectively. Notably, given each model pair, uncompressed methods need to write 24 (16 + 8) bits per model weight to disk, whereas \format significantly reduces the bit count 
to 10.7-11.5 (\cref{tbl:bits}), 
which is also smaller than the 19.1-19.6 bits incurred by separately compressing both models with \zstd. Expectedly, \format's number of incurred bits is in alignment with \format's high compression ratio (\cref{tbl:exp_size_both}).

\subsection{High Savings on Constrained Bandwidths}
\label{sec:exp_bandwidth}

This section studies the effect of I/O bandwidth on \format's time savings. We perform a parameter sweep on bandwidth from SSD by throttling with \texttt{systemd-run~\cite{systemd-run}} (verified using \texttt{iostat}~\cite{iostat}) and measure the time to load a model pair stored with \format vs uncompressed storage (\safetensors) at various bandwidths (\cref{fig:exp_bandwidth}). 

While \format is faster than uncompressed loading at all bandwidths, the speedup increases from $1.7\times$ (500MB/s) to $2.1\times$ and $2.2\times$ in the lowest bandwidth settings (20MB/s) for the small Llama 3.1 model and large Qwen 2.5 VL model, respectively.
Notably, the absolute time saving of \format versus uncompressed is 2483 seconds for loading the Qwen model at 20MB/s; this time saving significantly improves user experience in the common scenario where models are downloaded from cloud storage with limited network bandwidth (typically  30MB/s~\cite{zipnn}, grey vertical lines in \cref{fig:exp_bandwidth}).


\begin{figure}[t]\captionsetup[subfigure]{font=footnotesize}

\centering

\begin{subfigure}[b]{0.48\linewidth}
\begin{tikzpicture}

\begin{axis}[
    xtick=data,
    width=45mm,
    height=28mm,
    ymin=0,
    ymax=1000,
    axis y line*=none,
    axis x line*=none,
    xtick={1,2,3,4,5, 6,7,8},
    xticklabel style   = {align=center},
    ytick={0, 200, 400, 600, 800, 1000},
    yticklabels={0, 200, 400, 600, 800, 1000},
    xlabel=Read bandwidth (MB/s),
    xlabel style={yshift = 2.5ex},
    ylabel style={yshift=-4ex},
    xmin = 0,
    xmax = 500,
    xtick = {0, 100, 200, 300, 400, 500},
    xticklabels = {0, 100, 200, 300, 400, 500},
    tick label style={font=\scriptsize},
    legend style={
        at={(-0.2,1.1)},anchor=south west,column sep=2pt,
        draw=black,fill=white,
        /tikz/every even column/.append style={column sep=5pt},
        font=\scriptsize,
                inner ysep=0.1pt,
                legend image post style={xscale=0.6},
        font=\scriptsize
    },
    legend cell align={left},
    legend columns=4,
    label style={font=\scriptsize},
    ylabel={Load Time (s)},
    ymajorgrids,
]

\addplot[BaseBothColor, thick, mark = *, mark size=1pt]
table[x=x,y=y] {
x y
20 998.4
50 399.3
100 199.6
200 99.7
300 66.5
400 49.8
500 39.891
};
\addplot[OursColor, thick, mark = *, mark size=1pt]
table[x=x,y=y] {
x y
20 463
50 185
100 92.8
200 46.8
300 31.4
400 23.8
500 19.19
};
\addlegendentry{\safetensors}
\addlegendentry{\format \textbf{(Ours)}};
\draw[black, very thick, densely dashed,opacity=0.5] (axis cs: 30,0) -- (axis cs: 30, 1000);
\node[anchor=south west, align=left] at (axis cs: 30,600) {\scriptsize Cloud storage\\[-0.7em]\scriptsize  download speed};


\end{axis}
\end{tikzpicture}
\vspace{-6.5mm}
\caption{Llama 3.1 8B}
\end{subfigure}
\begin{subfigure}[b]{0.48\linewidth}
\begin{tikzpicture}

\begin{axis}[
    xtick=data,
    width=45mm,
    height=28mm,
    ymin=0,
    ymax=5000,
    axis y line*=none,
    axis x line*=none,
    xtick={1,2,3,4,5, 6},
    xticklabel style   = {align=center},
    xticklabels = {1600, 800, 400, 200, 100, 50},
    ytick={0, 1000, 2000, 3000, 4000, 5000},
    yticklabels={0, 1000, 2000, 3000, 4000, 5000},
    xlabel=I/O bandwidth (MB/s),
    xlabel style={yshift = 2.5ex},
    ylabel style={yshift=-4ex},
    xmin = 0,
    xmax = 600,
    xtick = {0, 100, 200, 300, 400, 500, 600},
    xticklabels = {0, 100, 200, 300, 400, 500, 600},
    tick label style={font=\scriptsize},
    legend style={
        at={(-0.2,1.1)},anchor=south west,column sep=2pt,inner ysep = 0.5pt,
        draw=black,fill=white,
        /tikz/every even column/.append style={column sep=5pt},
        font=\scriptsize,
    },
    legend cell align={left},
    legend columns=4,
    label style={font=\scriptsize},
    ylabel={Load Time (s)},
    ymajorgrids,
]

\addplot[BaseBothColor, thick, mark = *, mark size=1pt]
table[x=x,y=y] {
x y
20 4492.5
50 1795.6
100 898.487
200 449.461
300 299.593
400 229.309
500 186.09
600 157.35
};
\addplot[OursColor, thick, mark = *, mark size=1pt]
table[x=x,y=y] {
x y
20 2009.94
50 809.987
100 404.627
200 203.129
300 137.9
400 110.15
500 90.7
600 72.7
700 65.6
800 64.6
900 60.6
1000 58.1
1200 56.6
1500 56.4
};
\draw[black, very thick, densely dashed, opacity=0.5] (axis cs: 30,0) -- (axis cs: 30, 5000);
\node[anchor=south west, align=left] at (axis cs: 30,3000) {\scriptsize Cloud storage\\[-0.7em]\scriptsize  download speed};

\end{axis}
\end{tikzpicture}
\vspace{-6.5mm}
\caption{Qwen 2.5 VL 32B}
\end{subfigure}

\caption{\format's decoding time (secs) versus read bandwidth for two selected models. \format's smaller incurred storage size saves loading time by $2.2\times$ at lower bandwidths.
}
\label{fig:exp_bandwidth}
\end{figure}


\subsection{Faster Versus Online Quantization}
\label{sec:exp_online_quantization}
This section studies QStore's time savings for loading the low-precision model versus online quantization. We create QStore model pairs for the same high-precision model quantized with different quantization algorithms (GPT-Q and LLM.int8()), then measure time taken for \system to load the (i) low-precision model versus (ii) loading the high-precision model and then quantizing it on-the-fly.

We report results in \cref{fig:exp_online_quant}.
\system saves up to 47.3$\times$ loading time versus performing online quantization with the nonlinear, $2^{nd}$ order method GPT-Q (\cref{fig:exp_online_quant_gptq}), which performs complex and loss-aware quantization taking $\sim$7 minutes on the Mistral 7B model. Even when compared with the much faster quantization method LLM.int8() (as implemented in bitsandbytes), which only performs grouping, scaling and rounding in $<$ 2 seconds on Mistral 7B, \system still saves up to 2.53$\times$ in end-to-end model loading time due to \system's compressed model pair size being significantly smaller than the uncompressed high precision model (\cref{fig:exp_online_quant_bnb}).


\subsection{Handles Multi-Level Model Chains}
\label{sec:exp_multichain}
\begin{figure}[t]\captionsetup[subfigure]{font=footnotesize}
\pgfplotsset{scaled y ticks=false}
\centering
\begin{subfigure}[b]{0.48\linewidth}
\centering
\begin{tikzpicture}

\pgfplotstableread{
Label int conditional
OnlineQuantization 17.547 393.63
QStore 11.672 0.0
}\testdata

\begin{axis}[
    ybar stacked,
    clip=false,
    xlabel style={yshift = 1.5ex},
    xtick style ={draw=none},
    width=45mm,
    height=24mm,
    bar width=5mm,
    xtick={1, 2},
    xticklabels from table={\testdata}{Label},
    xmin=0.5,
    xmax=2.5,
    ylabel style={yshift = -5ex},
    axis y line*=none,
    axis x line*=none,
    ytick={0, 100, 200, 300, 400},
    yticklabels={0, 100, 200, 300, 400},
    y tick label style={yshift = 0ex},
    ymin=0,
    ymax = 450,
    ymajorgrids,
    tick label style={font=\scriptsize},
    legend style={
        font=\scriptsize,
        /tikz/every even column/.append style={column sep=0.5cm},
        legend columns = 4,
        at={(-0.28, 1.0)}, anchor=south west,
                legend image post style={xscale=0.6},
                        inner ysep=0.1pt,
        font=\scriptsize
    },
    label style={font=\scriptsize},
    ylabel={Time (s)},
    area legend
    ]

    \addplot [black,fill=BlueColor, postaction={
        pattern=grid
    }] table [y=int, meta=Label, x expr=\coordindex+1]{\testdata};
    \addlegendentry{Data Load}
    \addplot [black,fill=OursColor, postaction={
        pattern=crosshatch
    }] table [y=conditional, meta=Label, x expr=\coordindex+1] {\testdata};
    \addlegendentry[]{Quantization};

\end{axis}
    
    
\end{tikzpicture}
\vspace{-2.5mm}
\caption{GPT-Q, Qwen 2 7B}
\vspace{-1.5mm}
\label{fig:exp_online_quant_gptq}
\end{subfigure}
\begin{subfigure}[b]{0.48\linewidth}
\centering
\begin{tikzpicture}

\pgfplotstableread{
Label int conditional
OnlineQuantization 19.002 418.188
QStore 9.2422 0.0
}\testdata

\begin{axis}[
    ybar stacked,
    clip=false,
    xlabel style={yshift = 1.5ex},
    xtick style ={draw=none},
    width=45mm,
    height=24mm,
    bar width=5mm,
    xtick={1, 2},
    xticklabels from table={\testdata}{Label},
    xmin=0.5,
    xmax=2.5,
    ylabel style={yshift = -5ex},
    axis y line*=none,
    axis x line*=none,
    ytick={0, 100, 200, 300, 400},
    yticklabels={0, 100, 200, 300, 400},
    y tick label style={yshift = 0ex},
    ymin=0,
    ymax = 450,
    ymajorgrids,
    tick label style={font=\scriptsize},
    legend style={
        font=\scriptsize,
        /tikz/every even column/.append style={column sep=0.5cm},
        legend columns = 4,
        at={(-0.15, 1.1)}, anchor=south west
    },
    label style={font=\scriptsize},
    ylabel={Time (s)},
    area legend
    ]

    \addplot [black,fill=BlueColor, postaction={
        pattern=grid
    }] table [y=int, meta=Label, x expr=\coordindex+1]{\testdata};
    \addplot [black,fill=OursColor, postaction={
        pattern=crosshatch
    }] table [y=conditional, meta=Label, x expr=\coordindex+1] {\testdata};

\end{axis}
    
    
\end{tikzpicture}
\vspace{-2.5mm}
\caption{GPT-Q, Mistral 7B}
\vspace{-1.5mm}
\label{fig:exp_online_quant_gptq_second}
\end{subfigure}
\hfill
\begin{subfigure}[b]{0.48\linewidth}
\centering
\begin{tikzpicture}

\pgfplotstableread{
Label int conditional
OnlineQuantization 17.5476 1.98
QStore 10.8101 0.0
}\testdata

\begin{axis}[
    ybar stacked,
    clip=false,
    xlabel style={yshift = 1.5ex},
    xtick style ={draw=none},
    width=45mm,
    height=24mm,
    bar width=5mm,
    xtick={1, 2},
    xticklabels from table={\testdata}{Label},
    xmin=0.5,
    xmax=2.5,
    ylabel style={yshift = -5ex},
    axis y line*=none,
    axis x line*=none,
    ytick={0, 5, 10, 15, 20},
    yticklabels={0, 5, 10, 15, 20},
    y tick label style={yshift = 0ex},
    ymin=0,
    ymax = 20,
    ymajorgrids,
    tick label style={font=\scriptsize},
    legend style={
        font=\scriptsize,
        /tikz/every even column/.append style={column sep=0.5cm},
        legend columns = 4,
        at={(-0.05, 1.05)}, anchor=south west
    },
    label style={font=\scriptsize},
    ylabel={Time (s)},
    area legend
    ]

    \addplot [black,fill=BlueColor, postaction={
        pattern=grid
    }] table [y=int, meta=Label, x expr=\coordindex+1]{\testdata};
    \addplot [black,fill=OursColor, postaction={
        pattern=crosshatch
    }] table [y=conditional, meta=Label, x expr=\coordindex+1] {\testdata};

\end{axis}
    
    
\end{tikzpicture}
\vspace{-2.5mm}
\caption{LLM.int8(), Qwen 2 7B}
\label{fig:exp_online_quant_bnb}
\end{subfigure}
\begin{subfigure}[b]{0.48\linewidth}
\centering
\begin{tikzpicture}

\pgfplotstableread{
Label int conditional
OnlineQuantization 19.002 1.85
QStore 8.25 0.0
}\testdata

\begin{axis}[
    ybar stacked,
    clip=false,
    xlabel style={yshift = 1.5ex},
    xtick style ={draw=none},
    width=45mm,
    height=24mm,
    bar width=5mm,
    xtick={1, 2},
    xticklabels from table={\testdata}{Label},
    xmin=0.5,
    xmax=2.5,
    ylabel style={yshift = -5ex},
    axis y line*=none,
    axis x line*=none,
    ytick={0, 5, 10, 15, 20, 25},
    yticklabels={0, 5, 10, 15, 20, 25},
    y tick label style={yshift = 0ex},
    ymin=0,
    ymax = 25,
    ymajorgrids,
    tick label style={font=\scriptsize},
    legend style={
        font=\scriptsize,
        /tikz/every even column/.append style={column sep=0.5cm},
        legend columns = 4,
        at={(-0.05, 1.05)}, anchor=south west
    },
    label style={font=\scriptsize},
    ylabel={Time (s)},
    area legend
    ]

    \addplot [black,fill=BlueColor, postaction={
        pattern=grid
    }] table [y=int, meta=Label, x expr=\coordindex+1]{\testdata};
    \addplot [black,fill=OursColor, postaction={
        pattern=crosshatch
    }] table [y=conditional, meta=Label, x expr=\coordindex+1] {\testdata};

\end{axis}
    
    
\end{tikzpicture}
\vspace{-2.5mm}
\caption{LLM.int8(), Mistral 7B}
\label{fig:exp_online_quant_bnb2}
\end{subfigure}

\caption{\system's decoding via loading the low precision model saves up to $47.3\times$ time versus loading then quantizing the high precision model on-the-fly.}
\label{fig:exp_online_quant}
\vspace{-2mm}
\end{figure}
This section studies \system's generalizability to more-than-two model chains. We create a 3-model chain of FP16, INT8, and INT4 precisions using GPT-Q quantization with group size of 512, then compare storage cost of storing the model chain with \system (following procedures described in \cref{sec:implementation_sub}) versus alternative methods.

We report results in \cref{fig:exp_model_chain}. \system achieves up to 2.46$\times$ and 1.78$\times$ storage savings versus \safetensors and \zipnn+Zstd, respectively. Notably, \system's largest savings come from the FP16 and INT8 models: this is because \system stores the INT8 model as a conditional of the INT4 (i.e., INT8$|$INT4), then stores the FP16 model as a conditional of the INT8 (i.e., FP16$|$INT8). By bypassing the storage of redundant information (\cref{sec:experiment_space}), \system achieves up to 4.46$\times$ and 2.98$\times$ storage savings for the FP16 portion alone versus compression with \safetensors and \zipnn, respectively. Note that these savings are even greater than the model pair storage case because of the nature of our conditional encodings.


\begin{figure}[t]\captionsetup[subfigure]{font=footnotesize}
\pgfplotsset{scaled y ticks=false}
\centering
\begin{subfigure}[b]{0.48\linewidth}
\centering
\begin{tikzpicture}

\pgfplotstableread{
Label fp16 fp162 int8 int82 int4
Safetensors 5.25 0 2.625 0 1.3125
ZipNN+Zstd 3.508 0 2.408 0 1.073
QStore 0 1.176 0 1.475 1.013
}\testdata

\begin{axis}[
    ybar stacked,
    clip=false,
    xlabel style={yshift = 1.5ex},
    xtick style ={draw=none},
    width=48mm,
    height=26mm,
    bar width=5mm,
    xtick={1, 2, 3},
    xticklabels={Safetensors, ZipNN+Zstd, \system},
    xmin=0.5,
    xmax=3.5,
    ylabel style={yshift = -6ex},
    axis y line*=none,
    axis x line*=none,
    ytick={0, 2, 4, 6, 8, 10},
    yticklabels={0, 2, 4, 6, 8, 10},
    y tick label style={yshift = 0ex},
    ymin=0,
    ymax = 10,
    ymajorgrids,
    tick label style={font=\scriptsize, align=center},
    legend style={
        font=\scriptsize,
        /tikz/every even column/.append style={column sep=0.2cm},
        legend columns = 5,
        at={(-0.3, 1.1)}, anchor=south west,
                legend image post style={xscale=0.6},
                        inner ysep=0.1pt,
        font=\scriptsize
    },
    label style={font=\scriptsize},
    ylabel={Size (GB)},
    area legend
    ]

    \addplot [black,fill=BlueColor, postaction={
        pattern=grid
    }] table [y=fp16, meta=Label, x expr=\coordindex+1]{\testdata};
    \addlegendentry{FP16}
        \addplot [black,fill=BlueColor!50] table [y=fp162, meta=Label, x expr=\coordindex+1]{\testdata};
    \addlegendentry{FP16|INT8 (\system)}
    \addplot [black,fill=ZipnnColor, postaction={
        pattern=crosshatch
    }] table [y=int8, meta=Label, x expr=\coordindex+1] {\testdata};
    \addlegendentry[]{INT8};
        \addplot [black,fill=ZipnnColor!50] table [y=int82, meta=Label, x expr=\coordindex+1] {\testdata};
    \addlegendentry[]{INT8|INT4 (\system)};
        \addplot [black,fill=OursColor, postaction={
        pattern=grid
    }] table [y=int4, meta=Label, x expr=\coordindex+1] {\testdata};
    \addlegendentry[]{INT4};

\end{axis}
    
    
\end{tikzpicture}
\vspace{-6mm}
\caption{Llama 3.2 3B}
\label{fig:exp_model_chain_llama}
\end{subfigure}
\begin{subfigure}[b]{0.48\linewidth}
\centering
\begin{tikzpicture}

\pgfplotstableread{
Label fp16 fp162 int8 int82 int4
Safetensors 12.154 0 6.077 0 3.038
ZipNN+Zstd 8.182 0 5.529 0 2.441
QStore 0 2.840 0 3.434 2.441
}\testdata

\begin{axis}[
    ybar stacked,
    clip=false,
    xlabel style={yshift = 1.5ex},
    xtick style ={draw=none},
    width=48mm,
    height=26mm,
    bar width=5mm,
    xtick={1, 2, 3},
    xticklabels={Safetensors, ZipNN+Zstd, \system},
    xmin=0.5,
    xmax=3.5,
    ylabel style={yshift = -6ex},
    axis y line*=none,
    axis x line*=none,
    ytick={0, 5, 10, 15, 20, 25},
    yticklabels={0, 5, 10, 15, 20, 25},
    y tick label style={yshift = 0ex},
    ymin=0,
    ymax = 25,
    ymajorgrids,
    tick label style={font=\scriptsize, align=center},
    legend style={
        font=\scriptsize,
        /tikz/every even column/.append style={column sep=0.5cm},
        legend columns = 3,
        at={(-0.15, 1.1)}, anchor=south west
    },
    label style={font=\scriptsize},
    ylabel={Size (GB)},
    area legend
    ]

    \addplot [black,fill=BlueColor, postaction={
        pattern=grid
    }] table [y=fp16, meta=Label, x expr=\coordindex+1]{\testdata};
        \addplot [black,fill=BlueColor!50] table [y=fp162, meta=Label, x expr=\coordindex+1]{\testdata};
    \addplot [black,fill=ZipnnColor, postaction={
        pattern=crosshatch
    }] table [y=int8, meta=Label, x expr=\coordindex+1] {\testdata};
            \addplot [black,fill=ZipnnColor!50] table [y=int82, meta=Label, x expr=\coordindex+1] {\testdata};
        \addplot [black,fill=OursColor, postaction={
        pattern=grid
    }] table [y=int4, meta=Label, x expr=\coordindex+1] {\testdata};

\end{axis}
    
    
\end{tikzpicture}
\vspace{-2mm}
\caption{Qwen 2 7B}
\label{fig:exp_model_chain_qwen}
\end{subfigure}

\caption{\system's storage costs for a 3-model chain versus baselines. \system's conditional compression of high-precision models w.r.t. low-precision models saves up to 2.46$\times$ and 1.78$\times$ space versus uncompressed storage and \zipnn+Zstd.}
\label{fig:exp_model_chain}
\end{figure}
\section{Related Work}
\paragraph{Matrix Compression for Machine Learning}
Matrix compression has been extensively explored for ML model weights~\cite{elgohary2016compressed, bell2009implementing, kernert2014slacid, saad1990basic, li2019tuple}. These works exploit specific properties present in ML model weight matrices such as inherent sparsity, column correlations, and low distinct value counts~\cite{zhang2016materialization} to apply custom compression schemes more effective than off-the-shelf compression algorithms, aiming to fit the weight matrices in memory for performing efficient computations directly on the compressed data~\cite{saad1990basic, li2019tuple}. Grouping techniques are prevalent in these works, for example, CLA~\cite{elgohary2016compressed} jointly performs column grouping and selection of per-column-group compression algorithms to apply, SLACID~\cite{kernert2014slacid} and ~\cite{bell2009implementing} performs per-group compression on a matrix block granularity, and TOC~\cite{li2019tuple} applies compression on minibatches of input data.
In comparison, \system exploits the natural grouping incurred by quantization schemes to perform joint compression on multiple models of varying precision on LLM weights, which are more difficult to compress due to different properties such as inherent randomness and density~\cite{jaiswal2023emergence}.


\paragraph{Mixed-Strategy Compression}
Works have recently explored applying different compression algorithms on different data subsets to achieve higher overall compression rates~\cite{ilkhechi2020deepsqueeze, yu2020two, lance}. DeepSqueeze uses autoencoders to store tuples in tabular data, then applies per-dataset code-to-tuple mappings by brute-force evaluating and selecting the best-performing compression algorithm (e.g., run-length, delta compression). HIRE~\cite{yu2020two} uses reinforcement learning to estimate the best compression algorithm to apply to each data point in a timeseries.
While \system studies the different topic of multi-precision model storage, extending \system to incorporate techniques in these works such as replacing our currently-used Huffman compression on a per-model basis (\cref{sec:compression_encoding}) can be valuable future work.


\paragraph{Adaptive Data Formats for Machine Learning}
There exists works studying datatypes that adapt to the input data for ML model weights~\cite{mellempudi2017mixed, darvish2020pushing, courbariaux2014training, koster2017flexpoint, williamson1991dynamically}. These works have influenced many recent quantization schemes (e.g., LLM.int8()~\cite{llmint8dettmers}), focusing on dynamically selecting the exponent bit count for vectors on a per-group granularity based on data distribution (e.g., max/min values~\cite{williamson1991dynamically, courbariaux2014training}). Groups are often user-defined (e.g., in the form of tensors~\cite{williamson1991dynamically, courbariaux2014training} or group size $n$~\cite{darvish2020pushing}), while per-group exponent bit count selection can be done either reactively (e.g., increasing bits on overflow~\cite{williamson1991dynamically, courbariaux2014training}) or proactively (e.g., tracking value trend during model training~\cite{koster2017flexpoint}). \system complements these methods, being in principle applicable to these works' group-based techniques (now also seen in quantization) to reduce data sizes.
\begin{section}{Conclusion}
In this paper, we introduced \format, a unified file format for storing a high and low-precision model pair. \format defines a novel representation for storing the conditional information present in the high-precision model but not in the low-precision model. 
For model pair storage, \format stores the low-precision model, then applies novel grouping techniques on the conditional information to achieve efficient storage via high compression ratios.
Then, a model pair stored in the \format format can be losslessly decoded to load the low or high-precision model (or both).
We showed via experimentation that \format reduces the storage footprint of model pairs by up to $2.2\times$ while enabling up to $2\times$ and $1.6\times$ faster model pair saving and loading versus existing approaches, respectively, while also generalizing to a wide range of datatypes, quantization methods, and chains of more than two models.

\end{section}







\bibliographystyle{ACM-Reference-Format}
\bibliography{references}

\end{document}